\documentclass[12pt]{article}
\usepackage{graphicx}
\usepackage{natbib}
\usepackage{url} % not crucial - just used below for the URL 

\newcommand{\blind}{0}

\usepackage{dsfont}
\usepackage{amsmath,amsthm,amsfonts,amssymb}
\usepackage{hyperref}
\usepackage{algorithm}
\usepackage{algpseudocode}
\usepackage{kpfonts}
\usepackage{comment}
\usepackage{subcaption}
\usepackage{xcolor}
\usepackage{bbm}
\DeclareGraphicsExtensions{.png,.pdf}
\DeclareMathOperator*{\argmin}{arg\,min}
\usepackage{setspace}

\newcommand{\curly}[1]{\left\{#1\right\}}

\theoremstyle{definition}
\newtheorem{definition}{Definition}
\newtheorem{remark}{Remark}
\newtheorem{example}{Example}

\begin{document}

\def\spacingset#1{\renewcommand{\baselinestretch}%
{#1}\small\normalsize} \spacingset{2}

%%%%%%%%%%%%%%%%%%%%%%%%%%%%%%%%%%%%%%%%%%%%%%%%%%%%%%%%%%%%%%%%%%%%%%%%%%%%%%

\if0\blind
{
  \title{Conformal Prediction Sets for Populations of Graphs}
  \author{Anna Calissano \thanks{Corresponding author: a.calissano@imperial.ac.uk; We thank Safari Njema Polisocial Award 2018 for supporting the project.}\hspace{.2cm}\\
    Department of Mathematics, Politecnico di Milano\\
    now at Department of Mathematics, Imperial College London\\
    and \\
    Matteo Fontana \\
    Department of Mathematics, Politecnico di Milano\\
    now at Department of Computer Science, Royal Holloway University of London\\
    and\\
    Gianluca Zeni\\
    Department of Mathematics, Politecnico di Milano\\
    and\\
    Simone Vantini\\
    Department of Mathematics, Politecnico di Milano}
  \maketitle
} \fi

\if1\blind
{
  \bigskip
  \bigskip
  \bigskip
  \begin{center}
    {\LARGE\bf Conformal Prediction Sets for Populations of Graphs}
\end{center}
  \medskip
} \fi

\bigskip

\begin{abstract}
The analysis of data such as graphs has been gaining increasing attention in the past years. This is justified by the numerous applications in which they appear. Several methods are present to predict graphs, but much fewer to quantify the uncertainty of the prediction. The present work proposes an uncertainty quantification methodology for graphs, based on conformal prediction. The method works both for graphs with the same set of nodes (labelled graphs) and graphs with no clear correspondence between the set of nodes across the observed graphs (unlabelled graphs). The unlabelled case is dealt with the creation of prediction sets embedded in a quotient space. The proposed method does not rely on distributional assumptions, it achieves finite-sample validity, and it identifies interpretable prediction sets. To explore the features of this novel forecasting technique, we perform two simulation studies to show the methodology in both the labelled and the unlabelled case. We showcase the applicability of the method in analysing the performance of different teams during the FIFA 2018 football world championship via their player passing networks.
\end{abstract}

\noindent%
\vfill

%% \linenumbers

%% main text
\section{Introduction}
\label{sec:intro}

% OODA
One of the major challenges the discipline of statistics is facing in the latest years is how to deal with statistical units of increasing complexity. One of the first tasks in this endeavor is to find a proper mathematical embedding and to extend the standard elements of a statistician's toolbox (e.g clustering, principal component analysis, classification, regression). Indeed several papers on this line can be found in the literature. For example: functions embedded in a Euclidean space \citep{ramsay_when_1982,ramsay_functional_2005}, functions in non-Euclidean spaces \citep{menafoglio2016universal}, trees \citep{billera2001geometry,wang2007object,feragen2013toward}, graphs \citep{jain2009structure,durante2017nonparametric,chowdhury2019gromov,calissano2024populations} and in general, the whole Object-Oriented Data Analysis  framework \citep{marron_overview_2014,marron_object_2021}.

Despite numerous works in this stream of literature, uncertainty quantification has been relatively overlooked, regardless of its importance in application tasks (\cite{gneiting_probabilistic_2014}, Section 2.3.21 in \cite{petropoulos_forecasting_2022}). While some attempts at tackling this problem have been proposed in a functional setting \citep{degras_simultaneous_2011,antoniadis_prediction_2016} and for phylogenetic trees \citep{willis2019confidence}, to the best of our knowledge, uncertainty quantification for forecasting have not been proposed for more general data objects, let alone graph data.

In this paper, we focus our attention on the analysis of graph data, namely sets of different graphs generated by the same data-generating process. Two classical scenarios of populations of graphs are considered: populations of labelled and unlabelled graphs. Labelled graphs represent graphs where one can find a clear correspondence between the nodes across the observations. Unlabelled graphs instead represent those scenarios in which there is no clear correspondence between the nodes across the graphs. There are real world dataset of graphs which are naturally labelled (e.g., a dynamic network with a fix set of nodes), naturally unlabelled (e.g., graphs representing social interaction of different individuals) or partially labelled (e.g., graphs representing social interaction of different individuals who share a certain amount of friends). Several embeddings are available in the literature to describe unlabelled graphs \citep{chowdhury2019gromov, severn2022manifold, durante2017nonparametric}. In this work, we use the graph space proposed by \cite{jain2009structure} as a natural embedding for unlabelled graphs. Graph space is a quotient space, consisting of equivalence classes of permuted adjacency matrices. Intuitively, each element of this space is not a single graph, but the corresponding set of permuted graphs, obtained by the possible relabelling of the nodes of the original graph.

The aim of the present work is to develop a methodology to perform uncertainty quantification for populations of graphs embedded in a graph space. Due to the intrinsic complexity of this kind of data, we opt for the use of a model-free methodology for forecasting, namely the conformal prediction (CP) approach \citep[][for a unified review]{vovk_algorithmic_2023,fontana_conformal_2023}.
CP is a relatively novel method \citep{gammerman_learning_1998}, originally designed to provide uncertainty quantification for support vector machines. It relies on conformity measures. Given a cloud of points, conformity measures quantify how similar a data point is with respect to the members of this point cloud. Via very basic assumptions on the probability model behind the data (i.e. points need to be exchangeable) and on the non-conformity score (i.e. it needs to be invariant to permutations of the data), one is able to build prediction regions that come with good probabilistic properties. More specifically, CP regions display unconditional validity property not only asymptotically but also in finite samples. Another interesting feature of CP is that it does not specify much about the nature of the data. In fact, the approach requires just measurable spaces (see Chapter 2 in \cite{vovk_algorithmic_2023}), which do include the considered graph space.

The impact of the paper is twofold: we propose a framework to extend CP to general quotient spaces and we provide a tool for uncertainty quantification for graph case data. Alongside the theoretical interest, forecasting a new instance of a population of graphs has a clear and meaningful applicative interest in different fields, such as in the analysis of international trade networks \citep{amador_networks_2017}, Input-Output networks \citep{cerina_world_2015}, epidemic models where the subjects are connected using a graph topology \citep{ball_stochastic_2019}, brain connectivity of different subjects \cite{calissano2024graph}.

The paper is organized as follow. In Section \ref{sec:OODAnetwork}, we introduce in mathematical terms the graph space as a convenient embedding for graph data, both labelled and unlabelled. Section \ref{sec:labelled} is devoted to the definition of CP sets for populations of labelled graphs. In Section \ref{sec:unlabelled}, we generalize the framework to the unlabelled graphs case, extending CP sets to quotient metric spaces. In Section \ref{sec:Simulation}, we present two simulation studies. In the first one, we explore the performance of our method comparing different parametric and non-parametric prediction sets for labelled graphs. In the second one, we showcase the CP framework for a set of simulated unlabelled graphs. In Section \ref{sec:Application}, we analyse a set of player passing networks of different teams during the FIFA 2018 Football World Cup. We compute the CP sets for high-, low- and medium-performing teams, showcasing the potential of the framework in real-world applications. The whole framework is implemented as a module in the python package \texttt{geomstats} \citep{miolane2020geomstats}.

\section{The Mathematical Structure of a Population of Graphs}
\label{sec:OODAnetwork}
Let $\mathcal{G}=(V, E, a)$ be a graph with vertex set $V=\curly{v_1,\ldots,v_n}$, a set of edges $E\in V\times V$, and $a: E\rightarrow \mathbb{R}$ the attribute map. Each graph is represented as an adjacency matrix $x \in \mathbb{X},\, \mathbb{X}=\mathbb{R}^{n\times n}$, which is a standard representation for a graph. The entries $x(i,j)=a(i,j)$ of the matrix encode the edge weights when $i\neq j$ and the node weights when $i=j$. The distance between two adjacency matrices $x_1$ and $x_2$ is the sum of a distance function between the nodes and the edges attributes:
\begin{equation}
\label{eq:distance_total}
    d_{\mathbb{X}}(x_1,x_2)=\sum_{i=1}^{n}\sum_{j=1}^{n}{d(x_1(i,j),x_2(i,j))}
\end{equation}
where $d:\mathbb{R}\times \mathbb{R}\rightarrow \mathbb{R}$ is the chosen distance between the attributes. In our framework, $d$ is the squared Euclidean distance.

When moving from one to a set of graphs, a matching between the nodes is needed. If the nodes have a certain unique and non-interchangeable meaning across the graphs, the corresponding graphs are said to be \textit{labelled}. If instead, the nodes are interchangeable and their meaning is not unique but is related to the role they assume within the graph topology, the graphs are called \textit{unlabelled}. In other words, labelled graphs share the same sets of nodes while unlabelled graphs rely on a matching or aligned procedure of nodes to compare two graphs. Working with labelled or unlabelled graphs purely depends on the applicative context, the objective of the analysis, and the meaning of the nodes in the graph. Consider for example a player passing network in football, where the nodes represent the players of a given team and the edges the number of ball passages between players within a match (see Section \ref{sec:Application} for further details). If we compare two player passing networks of the same team, the networks are labelled as the nodes are the same. If the interest is in comparing two teams, the sets of players are different and the networks should be considered unlabelled, finding correspondence between players of different teams. Other examples are networks describing words co-occurrence in different novels \citep{severn2022manifold}, or brain connectivity of different patients \citep{simpson2013permutation, calissano2024graph}. The labelled and the unlabelled cases are the two extremes of a range of possibilities in comparing a set of nodes. Referring as the player passing networks as an example, the role of the players can be included in the analysis, working in a partially labelled situation - matching players of one team with players having the same role in another team. This work focuses on the unlabelled case as the most general one.

A natural representation for unlabelled graphs is expanding the graph from a single adjacency matrix to a set of permuted adjacency matrices (i.e., an equivalence class). This framework has been introduced by \cite{jain2009structure} for structured objects in general and studied in the particular case of graphs in \cite{calissano2024populations, guo2019quotient}, and \cite{kolaczyk2020averages}. Other possible representation for unlabelled graphs have been introduced in the literature \citep{severn2022manifold, maignant2023barycentric, chowdhury2019gromov, maignant2023barycentric}. However, these latter embeddings either relax the problem from discrete to continuous, making the interpretation of equivalence classes much harder, or consider a different isomorphism between graphs rather than the more natural permutation action.

In geometric terms, applying the permutation group action $\mathbb{T}$ to the space of adjacency matrices $\mathbb{X}$ by conjugation generates a quotient space $\mathbb{X}/\mathbb{T}$, called graph space. Every equivalence class $[x]=\{txt^T: t \in \mathbb{T}\}$ corresponds to all the graphs you can obtain from one adjacency matrix $x$ by permuting the nodes  (i.e., the rows and columns of the matrix $txt^T$, where $t \in \mathbb{T}$ is a permutation matrix of $\{0,1\}$ values). Given a $t\in \mathbb{T}$, we can also associate a unique function $\sigma^t:V\rightarrow V$ relabelling the nodes. In this geometrical framework, the labelled graphs are represented in the total space $\mathbb{X}$. 

\begin{figure}
    \centering
    \includegraphics[width=.5\textwidth]{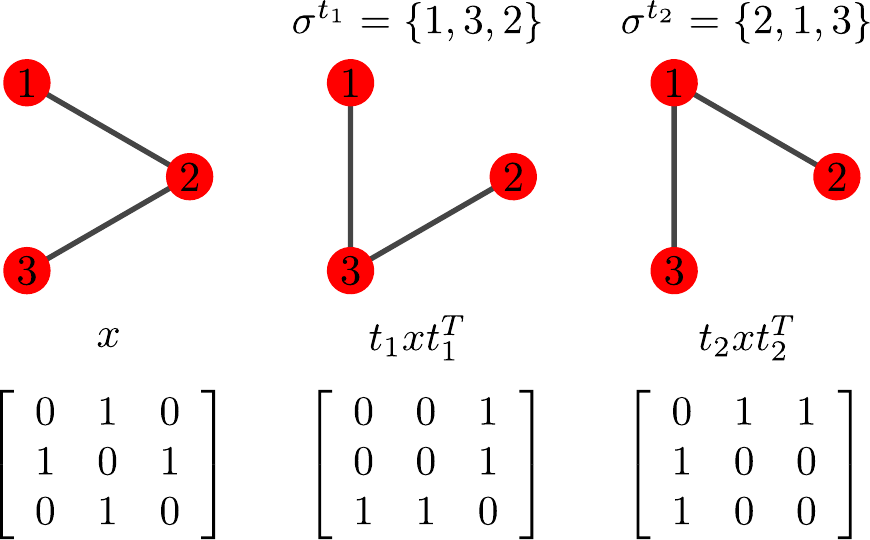}
    \caption{An example of an equivalence class of a graph $[x]$ with three nodes. We represent the graphs (top) and the corresponding adjacency matrices (bottom) to showcase the permutation action. $\sigma^{t_1}$ and $\sigma^{t_2}$ represent two permutations of $V$. $t_1, t_2$ represent the corresponding permutation matrices.}
    \label{fig:example_equivalence_class}
\end{figure}

In Figure \ref{fig:example_equivalence_class}, we show an example of equivalence class of a graph with three nodes $V=\{v_1,v_2,v_3\}$. The corresponding adjacency matrix is $x\in \mathbb{R}^{3\times 3}$. To show elements of the equivalence class, we consider two possible permutations $\sigma^{t_1}=\{1,3,2\}$ and $\sigma^{t_2}=\{2,1,3\}$, represented by two permutation matrices $T_1, T_2$.

From the definition of the distance in the $\mathbb{X}$ space, we can define the following distance in $\mathbb{X}/\mathbb{T}$:
\begin{equation}
\label{eq:distance_quotient}
        d_{\mathbb{X}/\mathbb{T}}([x_1],[x_2]])=\min_{t\in T}{d_\mathbb{X}(tx_1 t^T, x_2)}.
\end{equation}
This minimization problem is known as graph matching, which has been broadly studied in the literature (see for example \cite{bunke2000recent,conte2004thirty,livi2013graph} for an overview). When the $d_\mathbb{X}$ is the Frobenius norm, the state of the art suggests the Fast Quadratic Assignment algorithm \citep{vogelstein2015fast}. All the geometrical properties of $\mathbb{X}/\mathbb{T}$ are thoroughly described in \cite{calissano2024populations}. Here we report only the definitions we need for the purpose of this article. In particular, the projection function $\pi:\mathbb{X}\rightarrow \mathbb{X}/\mathbb{T}$ associates every element to its equivalent class $x\mapsto [x]=\{txt^T, \forall t\in T\}$.  The push-forward measure on $\mathbb{X}/\mathbb{T}$ is given by a measure on $\mathbb{X}$:

\begin{definition}
The graph space $\mathbb{X}/\mathbb{T}$ is endowed with a probability measure $\eta$ which is absolutely continuous with respect to the push-forward of the Lebesgue measure $m$ on $\mathbb{X}$. In particular, for $A \subset \mathbb{X}/\mathbb{T}$, we have $\eta(A) = 0$ if $m(\pi^{-1}(A)) = 0$. 
\end{definition}

\section{Prediction Parallelotopes for Labelled Graphs}\label{sec:labelled}

We start with the definition of CP for graphs in the labelled case. As we are working with adjacency matrices $x \in \mathbb{X}$ it is sufficient to adapt the general CP framework to work in such multidimensional Euclidean space $\mathbb{X}=\mathbb{R}^{n \times n}$. The labelled case serves as a starting point, preparatory to the more complex unlabelled case. 

Consider an i.i.d. population of graph adjacency matrices $\{ X_1, X_2,\ldots,X_k\}$, 
sampled from a distribution $\mathbb{P}$, and their realizations $\{ x_1, x_2,\ldots,x_k\}\subset \mathbb{X}$ . The problem we want to tackle is to quantify the uncertainty when predicting a new observation sampled from the same distribution. Formally, we want to define a prediction set $\mathcal{C}_{k,1-\alpha} := \mathcal{C}_{k,1-\alpha}(x_1,\ldots,x_k)$ such that
\begin{equation}
\label{eq_predictionset_labelled}
    \mathbb{P}\left( x_{k+1} \in \mathcal{C}_{k,1-\alpha} \right) \geq 1-\alpha,
\end{equation}
where $\alpha \in [0,1]$.

Differently from the univariate setting, the case of formulating prediction sets for complex data poses serious questions in terms of interpretability and practical usefulness of the obtained intervals. It is intuitive to understand that the best case in terms of interpretability for a prediction set is a region in space that allows a component-wise identification.

In more mathematical terms, we are interested in a set defined as:

\begin{equation}
\label{eq:eq_set}
\mathcal{C} := \left\{ x \in \mathbb{X}: x(i,j) \in \mathcal C(i,j) \quad \forall\, i, j = 1,\ldots,n \right\},
\end{equation}
where $\mathcal{C}(i,j) \subseteq \mathbb{R}$. The sets described in Equation \ref{eq:eq_set} are the Cartesian product of $p$ intervals of the real line. A set of this form is a parallelotope in $\mathbb{R}^{n\times n}$.

A prediction set in the shape of a parallelotope allows a practitioner to project the multivariate prediction region, which is valid at a level $\alpha$, in intervals for each element of $x$ without changing the coverage level. Why this specific shape is particularly desirable in the applications is thoroughly described in a series of papers in the case of functional data \citep{diquigiovanni_conformal_2021-1,diquigiovanni_distribution-free_2024,diquigiovanni_importance_2025}, where the authors argue how the interpretability of such regions is particularly natural. The same argument can be extended from functional to more complex data such as graphs, where having a prediction set for a edge or a node facilitates the interpretability. Our applied goal is to identify parallelotope-shaped sets with a given unconditional coverage level as in Equation \ref{eq:eq_set}. A method that has the explicit aim to identify prediction sets of the type described in Equation \ref{eq:eq_set} has shown to be CP \citep{diquigiovanni_importance_2025}. In fact, by aptly choosing a non-conformity score with the desired iso-contours and by focusing on Split/Inductive Conformal Framework (see  \cite{papadopoulos_inductive_2002} and \cite{lei_distribution-free_2014} for an introduction) one is able to obtain prediction sets with the desired shape. 
Our implementation of the CP framework for graph data begins by splitting the data  $\{x_1,\ldots,x_k\}$ into a training set $\{x_l\}, \, l \in \mathcal{I}_1$, and a calibration set $\{x_m\}, \, m \in \mathcal{I}_2$, where $\left|\mathcal{I}_1\right|+\left|\mathcal{I}_2\right|=k$ and $\mathcal{I}_1\cup \mathcal{I}_2 = \{1, \dots, k\}$ .
We then train an algorithm $\mathcal{A}$ on the training set:
\begin{equation}
   \hat{\mu} = \mathcal{A}\left( \left\{x_l, l\in\mathcal{I}_1 \right\} \right).
\end{equation}
$\mathcal{A}$ is a generic algorithm (e.g., Sturm algorithm for the estimation of the Fréchet mean \cite{sturm}). There are no specific theoretical requirements with respect to the choice of $\mathcal{A}$, apart from being a symmetric function of the data points.
We can use $\hat{\mu}$ to compute a set of non-conformity scores as
\begin{equation}
\label{eq_NCM}
  R_{m} = \max_{i,j\in{1,\dots,n}} \left| x_m(i,j) - \hat{\mu}(i,j) \right| , \quad m\in \mathcal{I}_2. 
\end{equation}
In other words, we use as the non-conformity score the absolute value of the elements of the residual matrix $x_m - \hat{\mu}$. This non-conformity score allows us to define an empirical p-value as
\begin{equation*}
 \forall x_m,\, m\in \mathcal{I}_2, \quad p_{x_m} :=  \frac{\left|\left\{i \in  \mathcal{I}_2  : R_{i} \geq R_{m}\right\}\right|}{\left|\mathcal{I}_2\right|+1}
\end{equation*}
and a conformal prediction region of a given $\alpha$ as
\begin{equation}
\label{eq_predictionset_1}
\mathcal{C}_{k,1-\alpha}:=\left\{x \in \mathbb{X}: p_x > \alpha \right\}.    
\end{equation}

Starting from  the non-conformity score (Equation \ref{eq_NCM}) and the definition of the conformal prediction region (Equation \ref{eq_predictionset_1}), one can say that $x_{k+1}\in \mathcal{C}_{k, 1-\alpha} \iff R_{k+1} \leq h$, with $h$ the $\lceil (\left| \mathcal{I}_2\right|+1)(1-\alpha) \rceil$-th smallest value in the set $\{ R_m: m \in \mathcal{I}_2 \}$. 
Then, the split CP set induced by the non-conformity score can be reformulated as:
\begin{equation}
\mathcal{C}_{k, 1-\alpha}:= \left\{x \in \mathbb{X}: x(i,j)\in [\hat{\mu}(i,j)-h, \hat{\mu}(i,j)+h] \quad \forall i,j=1,\dots, n \right\},
\label{eq_pred_set_max}
\end{equation}
by rewriting  Equation \ref{eq_NCM} as:
%\begin{align}
%\max_{i,j\in{1,\dots,n}} \left| x_{k+1}(i,j)-\hat{\mu}(i,j)\right| \leq h & \quad  \Rightarrow
%\left| x_{k+1}(i,j)-\hat{\mu}(i,j)\right| \leq h  \quad \forall\ i,j \quad
% \Rightarrow \\ \Rightarrow x_{k+1}(i,j) &\in [\hat{\mu}(i,j)-h, \hat{\mu}(i,j)+h] \quad \forall i,j.
% \end{align}

$$
\max_{i,j\in{1,\dots,n}} \left| x_{k+1}(i,j)-\hat{\mu}(i,j)\right| \leq h  \quad 
 \Rightarrow x_{k+1}(i,j) \in [\hat{\mu}(i,j)-h, \hat{\mu}(i,j)+h] \quad \forall i,j.$$
The calculation of these sets is very convenient: we require to train the central tendency estimation algorithm $\mathcal{A}$ only once, and we have a closed form to identify the set.

\subsection{Length Modulation}
\label{sec:amplitudemodulation_labelled}
The main shortcoming of the approach proposed above is that the identified parallelotope has constant length across all $i,j=1,\dots,n$. Each projection of the prediction set $\mathcal{C}(i,j)$ over the attribute identified by the tuple $(i,j)$ is of constant length $2h$. While there may be situations in which such a feature is desirable, practitioners usually face cases where edge attributes have different variability, and may want to take this into account when making a global prediction (for instance, with wider or narrower sets). The extreme situation happens when a vertex or an edge has constant attribute or it is completely absent from the population of graphs analysed, meaning that $x(i,j)=0$. Any length different from $0$ for the interval in $\mathcal{C}_{k, 1-\alpha}(i,j)$ is not desirable.

Following \cite{lei_distribution-free_2018}, we condition the length of Equation  
 b \ref{eq_pred_set_max} across $i,j \in \{ 1,\ldots,n \}$ using a local notion of variability. Namely, we modify Equation \ref{eq_NCM} in the following fashion:

\begin{equation}
\label{eq_modulated_general}
  R_{m} = \max_{i,j\in {1,\dots,n}} \left[ \frac{ \left|x_m(i,j) - \hat{\mu}(i,j) \right| }{\hat{s}(i,j)} \right], 
\end{equation}
where $\hat{s}=\mathcal{S}\left( \left\{x_l,l\in \mathcal{I}_1 \right\} \right)$ is an estimator of local variability, trained on the set $\left\{x_l\right\}, \,l\in \mathcal{I}_1 $ using an algorithm $\mathcal{S}$. As it was the case with $\mathcal{A}$, it is important that $\mathcal{S}$ is computed only on the training set, and it is a symmetric function of the training set ($\hat{s}$ should not change regardless of a permutation of points in my training set).
\begin{comment}

\begin{algorithm}
\caption{Split CP Parallelotopes for Populations of Graphs with Length Modulation}
\label{algo_split_mod}
\begin{algorithmic}[1]
\State \algorithmicrequire\, Data $\{x_i,\,i=1,\ldots,k\}$, type-1 error level $\alpha\in(0,1)$, regression algorithm $\mathcal{A}$, length modulation algorithm $\mathcal{S}$

\State split randomly $\left\{ 1,\ldots,k\right\}$ into two subsets $\mathcal{I}_1,\mathcal{I}_2$
\State $\hat{{\mu}}= \mathcal{A}\left(\left\{ x_l,\,l\in\mathcal{I}_1\right\}\right)$, $\hat{s}=\mathcal{S}\left(\left\{ x_l,\,l\in\mathcal{I}_1\right\}\right) $

\State $R_{m} = \:  \max_{i,j} \left( \frac{\left| x_m(i,j) - \hat{\mu}(i,j)\right|}{\hat{s}(i,j)} \right),\: m\in\mathcal{I}_2$
\State $h =$ is the $\lceil (\left| \mathcal{I}_2\right|+1)(1-\alpha) \rceil$-th smallest value in the set $\{ R_m: m \in \mathcal{I}_2 \}$

\State \algorithmicensure\, $\mathcal{C}_{k, 1-\alpha}:= \left\{x \in \mathbb{X}: x(i,j) \in [\hat{\mu}(i,j)-h\hat{s}(i,j), \hat{\mu}(i,j)+h\hat{s}(i,j)] \quad \forall i,j=1,\dots,n \right\}$
\end{algorithmic} 
\end{algorithm}
\end{comment}
According to the choice of the algorithm $\mathcal{S}$, the modulating behaviour dramatically changes. We mention two notable cases. When $\hat{s}(i,j)=1 \, \forall i,j $ and no modulation is taking place, we get sets composed of intervals of constant length, as defined in Equation \ref{eq_pred_set_max}. When $\hat{s}(i,j)=\sqrt{Var(x(i,j)}$, where $Var(\cdot)$ is the sample variance, the resulting set length will be modulated according to the local variability of the attributes of the graph. Notice that a small perturbation $\hat{s}(i,j)=\sqrt{Var(x(i,j)}+\epsilon$ is added to avoid indeterminate form when $x(i,j)$ is constant across all population.

\section{Prediction Class of Parallelotopes for Unlabelled Graphs}\label{sec:unlabelled}
We now extend the concepts introduced in the previous section to the more general unlabelled framework (see Section \ref{sec:OODAnetwork} for a detailed introduction). Extending the setting to unlabelled graphs is only possible thanks to the specific construction of graph space based on discrete group action applied to the space of adjacency matrices $\mathbb{R}^{n\times n}$. Firstly, we look at how to  extend a parallelotope to the quotient space.

\begin{definition}
A set $\mathcal{C}\in \mathbb{X}$ as defined in Equation \ref{eq:eq_set} can be projected on the graph space as: 
\begin{equation*}
    [\mathcal{C}]=\bigcup_{t=1}^{|T|}\varprod_{i,j=1}^{n}{\mathcal{C}(\sigma^t(i),\sigma^t(j))}, \quad [\mathcal{C}]\subseteq \mathbb{X}/T,
\end{equation*} where $\sigma^t$ is the nodes relabelling function associated to the permutation $t\in \mathbb{T}$.
\end{definition}
The idea is to define a set of intervals that follow the index permutation of the elements in the graph space. For the sake of simplicity, we define:
$$\mathcal{C}^t=\varprod_{i,j=1}^{n}{\mathcal{C}(\sigma^t(i),\sigma^t(j))}$$
where $\mathcal{C}(\sigma^t(i),\sigma^t(j)) \subseteq \mathbb{R}$, $\mathcal{C}^t\subseteq\mathbb{R}^{n\times n}$.

The probability of this interval in the graph space can be computed using the projection on the total space $X$: 
 $$\mathbb {P}_\eta \left(\bigcup_{t=1}^{|T|}\mathcal{C}^t\right)=\sum _{t=1}^{|T|}\left((-1)^{t-1}\sum _{I\subseteq \{1,\ldots ,|T|\} \atop |I|=t}\mathbb {P} (A_{I})\right),$$
where 
$A_{I}:=\bigcap _{{t\in I}}\mathcal{C}^t$.

\begin{example}

\begin{figure}[hbpt!]
    \centering
       \includegraphics[width=\textwidth]{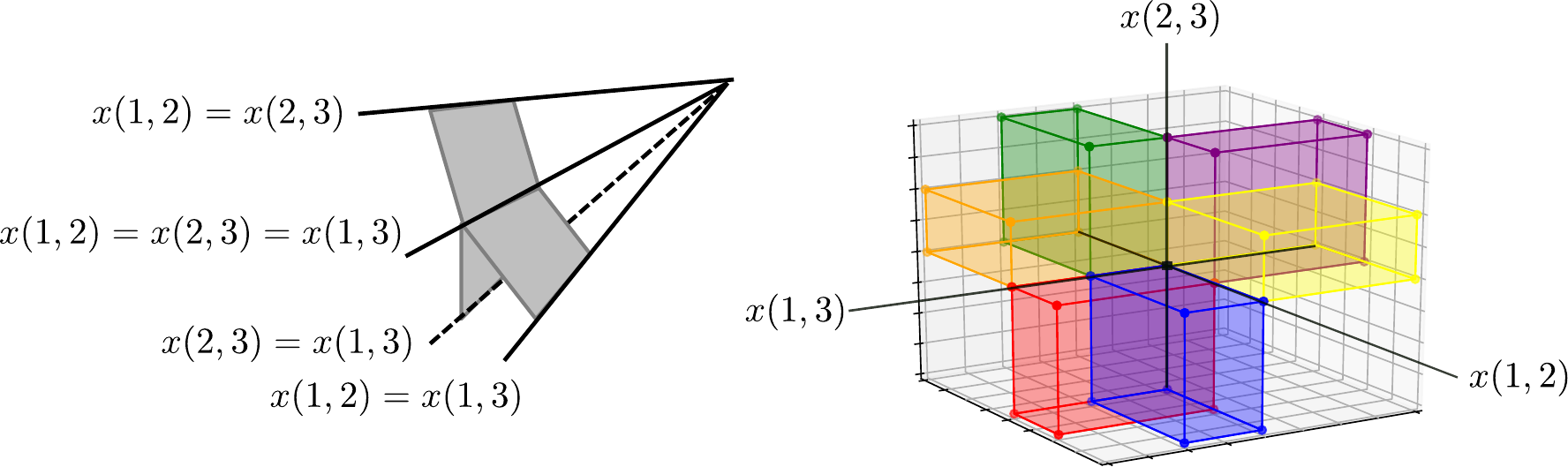}
        \caption{Consider an undirected graph with three nodes and with attributes on edges only. It can be represented by three matrix entries: $x(1,2), x(1,3), x(2,3)$. Right: Conceptual visualization of the shape of a set for the graph in the graph space. Left: its back-projection $\pi^{-1}$ in the total space.}
    \label{fig:interval}
\end{figure}
Consider the undirected graph with $n=3$ three nodes represented in Figure \ref{fig:example_equivalence_class}. To visualize the conformal prediction set, we constrain the graph to be undirected - meaning $x(i,j)=x(j,i)$ for all $i,j$ - and with no attributes on nodes - $x(i,i)=0$ for all $i$. Then, the only non-null entries of the adjacency matrix are $x(1,2), x(2,3), x(1,3)$: the graph is a point in $\mathbb{R}^3$. The number of permutations is $3!=6$. The interval $\mathcal{C}$ is the Cartesian product of three intervals on the real line $\mathcal{C}=\mathcal{C}(1,2)\times \mathcal{C}(2,3)\times \mathcal{C}(1,3)$. If we permute this shape with the following permutation $\sigma^{t_1}=\{2,1,3\}$, we obtain a new set $\mathcal{C}^t=\mathcal{C}(\sigma^t(1),\sigma^t(2))\times \mathcal{C}(\sigma^t(2),\sigma^t(3))\times \mathcal{C}(\sigma^t(1),\sigma^t(3))=\mathcal{C}(2,1)\times \mathcal{C}(1,3)\times \mathcal{C}(2,3)$. In Figure \ref{fig:interval}, we represent the whole set in the quotient space - in this case reduced to $\mathbb{R}^3$ up to permutations - and its projection on the total space, as the union of all the possible permutations.
\end{example}

As in the labelled case, we can define the interval with a given coverage level:

\begin{equation}
\mathbb{P_\eta}\left( [x_{k+1}] \in [\mathcal{C}]_{k,1-\alpha} \right) \geq 1-\alpha
\end{equation}
 We start by splitting our set of unlabelled graphs $\{[x_1],\ldots,[x_k]\}$ in a training set $\mathcal{I}_1$ and a calibration set $ \mathcal{I}_2$, where $\left|\mathcal{I}_1\right|+\left|\mathcal{I}_2\right|=k$ and $\mathcal{I}_1 \cup \mathcal{I}_2 = \{1, \dots, k\}$.

One can compute an empirical p-value defined exactly as in the labelled case:

\begin{equation*}
\quad p_{[x_m]} :=  \frac{\left|\left\{i \in  \mathcal{I}_2  : R_{i} \geq R_m\right\}\right|}{\left|\mathcal{I}_2\right|+1} \quad \forall [x_m],\, m\in \mathcal{I}_2
\end{equation*}
where $R$ is a non-conformity score as defined by \cite{vovk_algorithmic_2023}. The CP set defined using the above definition of p-value can be identified as
\begin{equation}
\label{eq_predictionset}
[\mathcal{C}]_{k,1-\alpha}:=\left\{[x] \in \mathbb{X}/T: p_{[x]} > \alpha \right\},
\end{equation}
given a generic algorithm $\hat{\mu} = \mathcal{A}\left( \left\{[x_l], l\in\mathcal{I}_1 \right\} \right)$. We define our non-conformity score $R_{m}$, $\forall m \in \mathcal{I}_2$ to be
\begin{equation}
\label{eq:ncm_unlabelled}
  R_{m}=\max_{i,j=1,\dots,n} \left|x_m(\sigma^{t_m}(i),\sigma^{t_m}(j)) - \hat{\mu}(i,j)\right|,\quad \forall m \in \mathcal{I}_2
\end{equation}

where $\sigma^{t_m}$ is the permutation map associated to the permutation matrix\\
$t_m = \argmin_{t\in T}(d_{\mathbb{X}}(tx_m t^T,\hat{\mu}))$.
This non-conformity score selects the permutation that optimally aligns the graph with the estimator and consequently selects edges or nodes that are most far apart from each other.

In the case of length modulation, the Equation \ref{eq:ncm_unlabelled} becomes:

\begin{equation}
\label{eq:ncm_unlabelled_mod}
  R_m=\max_{i,j=1,\dots,n}\left[\frac{ \left|x_m(\sigma^{t_m}(i),\sigma^{t_m}(j)) - \hat{\mu}(i,j)\right|}{\hat{s}(i,j)}\right]
\end{equation}
where $\hat{s}=\mathcal{S}(\{[x_l], l \in \mathcal{I}_1\})$ is an estimator of the variability of the edge or node computed on the training set. We compute the component wise variability of nodes and edges after the alignment with respect to the central estimator $\hat{\mu}$.  The whole procedure is described in Algorithm \ref{algo_unlabel}.

\begin{algorithm}
\caption{Split conformal prediction parallelotopes for populations of unlabelled graphs with length modulation}
\label{algo_unlabel}
\begin{algorithmic}[1]
\State \algorithmicrequire\, Data $\{[x_i],\,i=1,\ldots,k\}$, type-1 error level $\alpha\in(0,1)$, predictive algorithm $\mathcal{A}$, length modulation algorithm $\mathcal{S}$
\State split randomly $\left\{ 1,\ldots,k\right\}$ into two subsets $\mathcal{I}_1,\mathcal{I}_2$
\State Compute the estimator $\hat{\mu}= \mathcal{A}\left(\left\{ [x_l],\,l\in\mathcal{I}_1\right\}\right)$
\State Find $\{t_1,\dots,t_{|\mathcal{I}_1|}\}$ s.t. $  t_l = \argmin_{t\in T}(d_{x}(t x_l t^T,\hat{\mu})) $
\State Compute the length modulation using the aligned graphs  $\hat{s}=\mathcal{S}\left(\left\{ t_l x_l t_{l}^{T},\,l\in\mathcal{I}_1\right\}\right) $
\State Find $\{t_1,\dots,t_{|\mathcal{I}_2|}\}$ s.t. $  t_m = \argmin_{t\in T}(d_{x}(t x_m t^T,\hat{\mu})) $
\State $R_m = \: \max_{i,j=1,\dots, n} \left( \frac{|(t_m x_m t_m ^T)(i,j) - \hat{\mu}(i,j)|}{\hat{s}(i,j)} \right)=  \max_{i,j=1,\dots, n} \left( \frac{| x_m((\sigma^{t_m}(i),\sigma^{t_m}(j)) - \hat{\mu}(i,j)|}{\hat{s}(i,j)} \right),\: m\in\mathcal{I}_2$
\State $h$ is equal to $\lceil (\left| \mathcal{I}_2\right|+1)(1-\alpha) \rceil$-th smallest value in the set $\{ R_m: m \in \mathcal{I}_2 \}$

\State \algorithmicensure\, $\mathcal{C}_{k, 1-\alpha}:= \left\{ [x] \in \mathbb{X}/T: (tx)(i,j) \in [\hat{\mu}(i,j)-h\hat{s}(i,j), \hat{\mu}(i,j)+h\hat{s}(i,j)] \forall j=1,\dots,p, \forall t  \in T \right\}$
\end{algorithmic} 
\end{algorithm}

\begin{remark}
The reader should note how it is not required to specify anything about $\mathcal{A}$. This generality, which is one of the main interesting features of the CP framework, allows for the use of any predictive algorithm, either statistically inspired or machine-learning inspired. For explanatory purposes, we are going to use the Fréchet mean as the $\mathcal{A}$ (see \cite{calissano2024populations} for definitions and details). However, other methods can be easily tested thanks to the flexible implementation in \textsc{geomstats} \citep{miolane2020geomstats}.
\end{remark}

\section{Simulation studies}
\label{sec:Simulation}
We illustrate the theoretical results described in the previous section on two simulated datasets and one case study. 
In all these examples, the function $\mathcal{A}$ is going to be a Fréchet mean estimator. In the labelled case, the Fréchet mean corresponds to the arithmetic sample mean, while in the unlabelled case, the Fréchet mean is computed with the Align All and Compute Procedure (see Algorithm 1 in \cite{calissano2024populations} for further details). The CP Parallelotopes is implemented as the module \texttt{Graph Space} within the \texttt{geomstats} python package \citep{miolane2020geomstats}.

\subsection{Simulation: Labelled Case}
We compute the Empirical Coverage of different parametric intervals and the CP intervals. We generate a set of $130$ graphs ($|\mathcal{I}_1|+|\mathcal{I}_2|=130$, where $|\mathcal{I}_1|=30$ is the training set and $|\mathcal{I}_2|=100$ is the calibration set). Every directed graph has $5$ nodes with Gaussian attributes $x(i,i) \sim N(0,1)$ and $20$ edges $x(i,j)$ following three different distributions: (1) Gaussian attributes $N(0,1)$; (2) Uniform attributes $U(-1.7,1.7)$; (3) t-Student attributes with $1$ degree of freedom. Having two different distributions on node and edge attributes is very common in applications because nodes and edges usually describe two different phenomena. For every generated model, we compute the sample mean and two different prediction intervals: (1)  Univariate Gaussian Intervals with Bonferroni Correction $\hat{x}(i,j)\pm t_{k-1}((\alpha/n^2)/2)\sqrt{1+\frac{1}{k}}\hat{s}(i,j)$; (2) Simultaneous Gaussian Intervals: $\bar{x}(i,j)\pm \hat{s}(i,j)\sqrt{\left(1+\frac{1}{k}\right)\frac{(k-1)p}{(k-p)}F_{(p,k-p)}(\alpha)}$, 
where $t_{k-1}((\alpha/n^2)/2)$ and  $F_{(p,k-p)}(\alpha)$ denotes the upper quantile.  $\hat{s}_i$ is the estimated sample variance and $\hat{x}$ the sample mean.

\begin{figure}
    \centering
    \includegraphics[width=\textwidth]{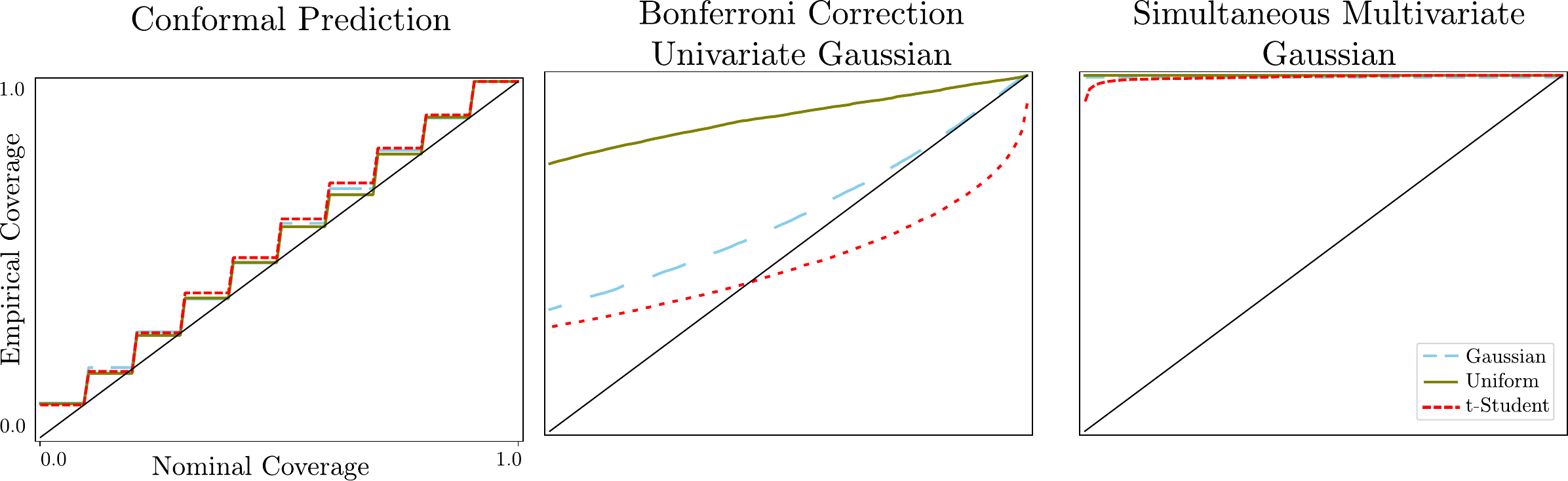}
    \caption{ Empirical coverage as a function of nominal coverage $\alpha$. Each plot corresponds to a different interval. Each line in the plots corresponds to a different generative model. The diagonal represents the theoretical quantiles.}
    \label{fig:Empirical_Coverage}
\end{figure}

We run every experiment $r=\{1, \dots, 100\}$ times. For $100$ $\alpha$ considered, we compute the empirical coverage on the test set, defined as:
\begin{equation}
    \hat{E}(1-\alpha)=\frac{1}{100}\sum_{k=1}^{130}{\sum_{r=1}^{100}\frac{{\mathbbm{1} x_{k,r}{\in \mathcal{C}_{k,1-\alpha,r}}}}{100}}.
\end{equation}
where the tuple $(k,r)$ refers to graph $k$ of experimental run $r$.
In Figure \ref{fig:Empirical_Coverage}, we show the calibration curves $\alpha$ for the different generative models and the different intervals and we compare them with the theoretical quantiles.
Bonferroni-corrected intervals are quite conservative in the Gaussian case and very conservative in the uniform one. With respect to the t-Student, we observe a generic conservativeness for low levels of nominal coverage, generated by the Bonferroni correction: this effect tends to disappear for higher coverage levels. We see that Bonferroni-corrected intervals are under-covering for nominal levels that are commonly used in the practice. The projection over the components of multivariate Gaussian intervals generates, similarly to our method, prediction sets with the shape of a parallelotope: it appears evident how they are conservative. The proposed conformal method is unique in its ability to provide properly calibrated prediction sets, regardless of the distribution.

\subsection{Simulation: Unlabelled Case}
\begin{figure}[!htbp]
\centering
\includegraphics[width=\textwidth]{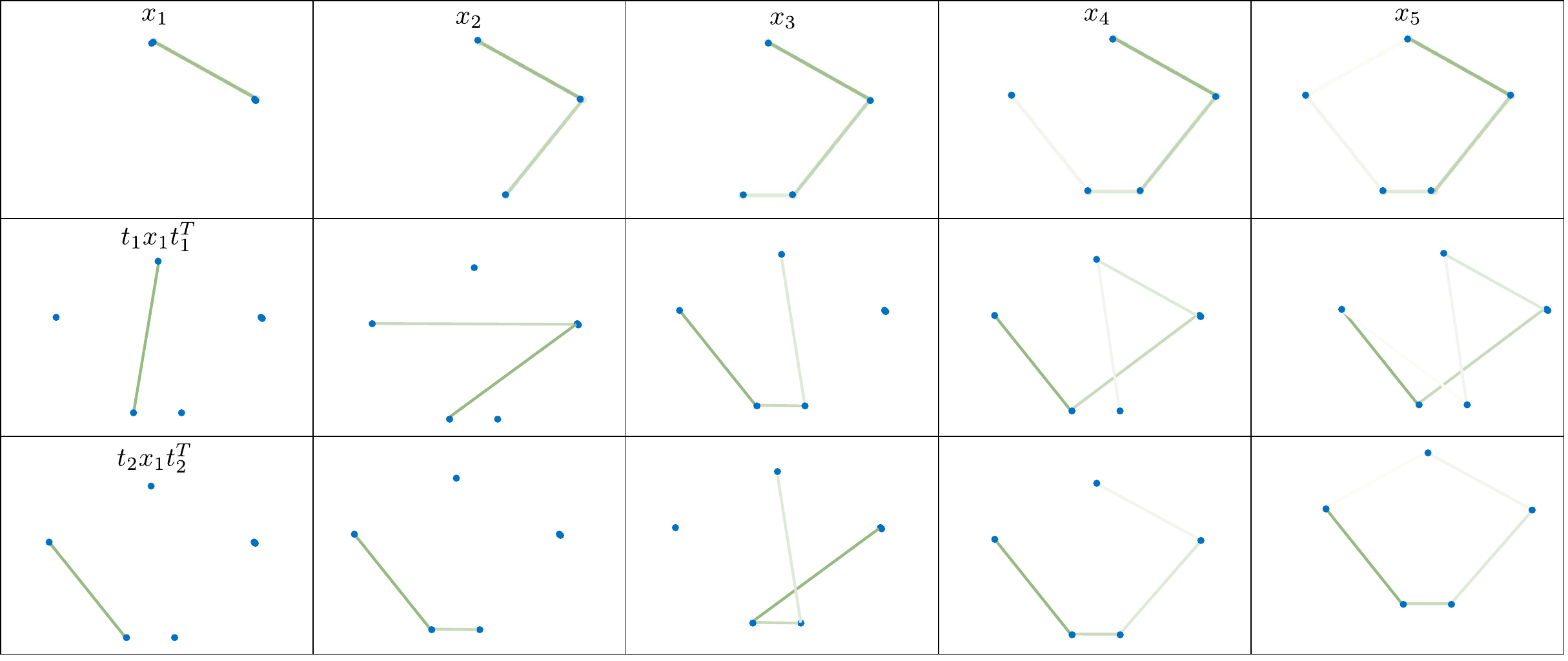} 
\caption{Example of unlabelled dataset: each column is an equivalence class of graphs with a given topology. Each row shows examples of elements in each equivalence class. While the nodes have fixed attributes, the color of the edge is proportional to the intensity of the edge attribute.}
\label{fig:labeled_pentagons}
\end{figure}

In this example, we simulate $500$ graphs from the equivalence classes $\{[x_1],\dots , [x_5]\}$ shown in Figure \ref{fig:labeled_pentagons}. The graphs have constant attributes on nodes $x(i,i) = 10$ and decreasing attributes on edges $x(i,j)\in \{20,40,60,80,100\}$, when $i\neq j$. The edge values are displayed by the color gradient of the edges in Figure \ref{fig:labeled_pentagons}.

\begin{figure}[!htbp]
\centering
\begin{subfigure}{0.3\textwidth}
\includegraphics[width=\linewidth]{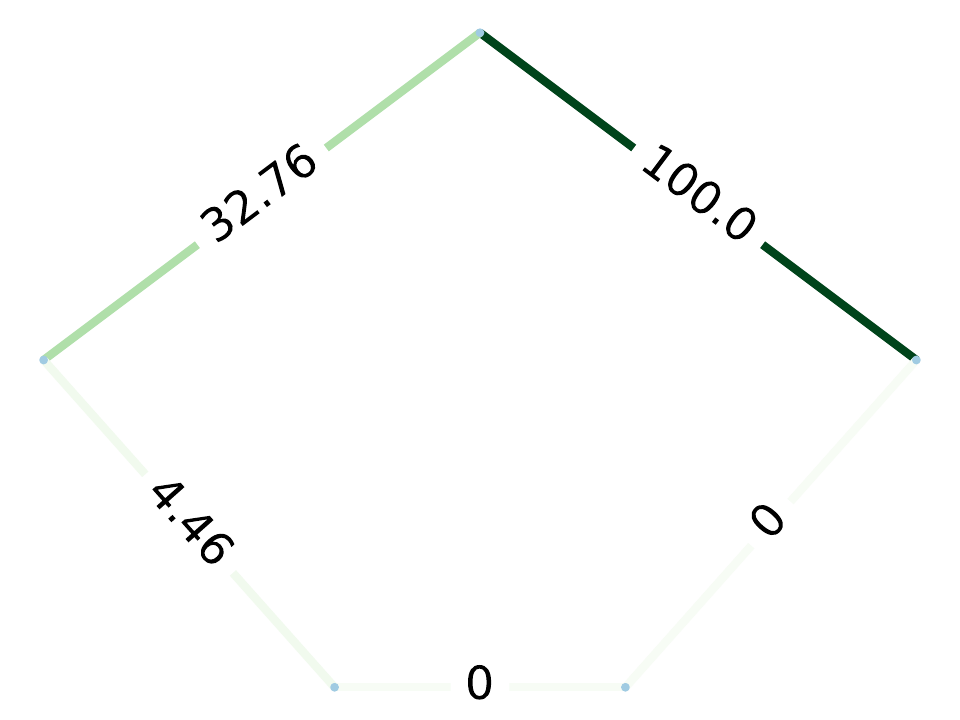} 
\caption{Minimum}
\end{subfigure}
\begin{subfigure}{0.3\textwidth}
\includegraphics[width=\linewidth]{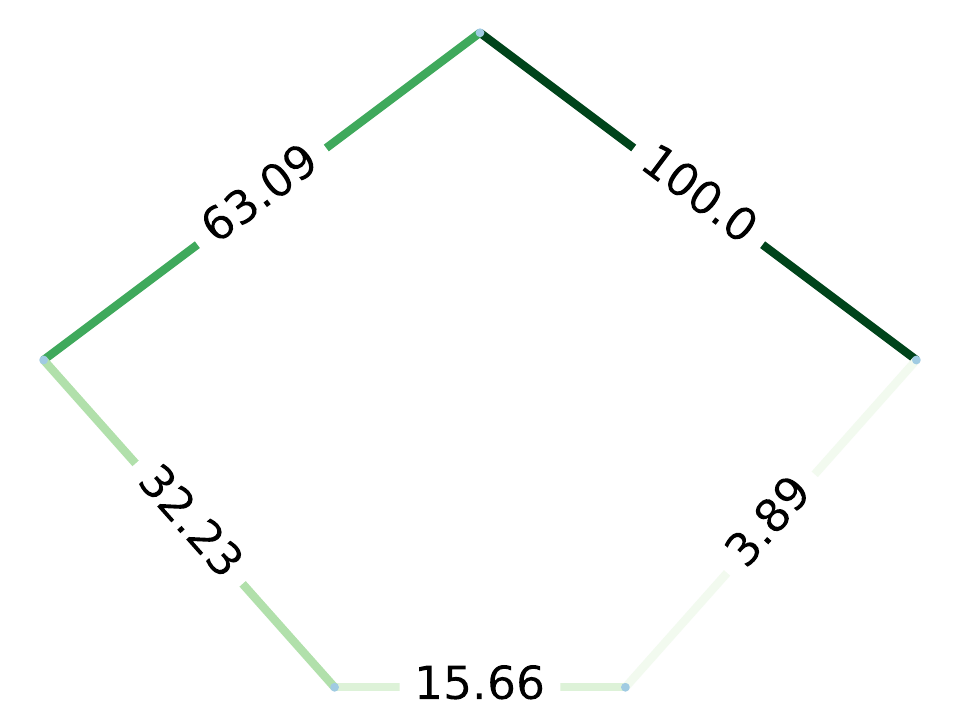} 
\caption{Fréchet mean}
\end{subfigure}
\begin{subfigure}{0.3\textwidth}
\includegraphics[width=\linewidth]{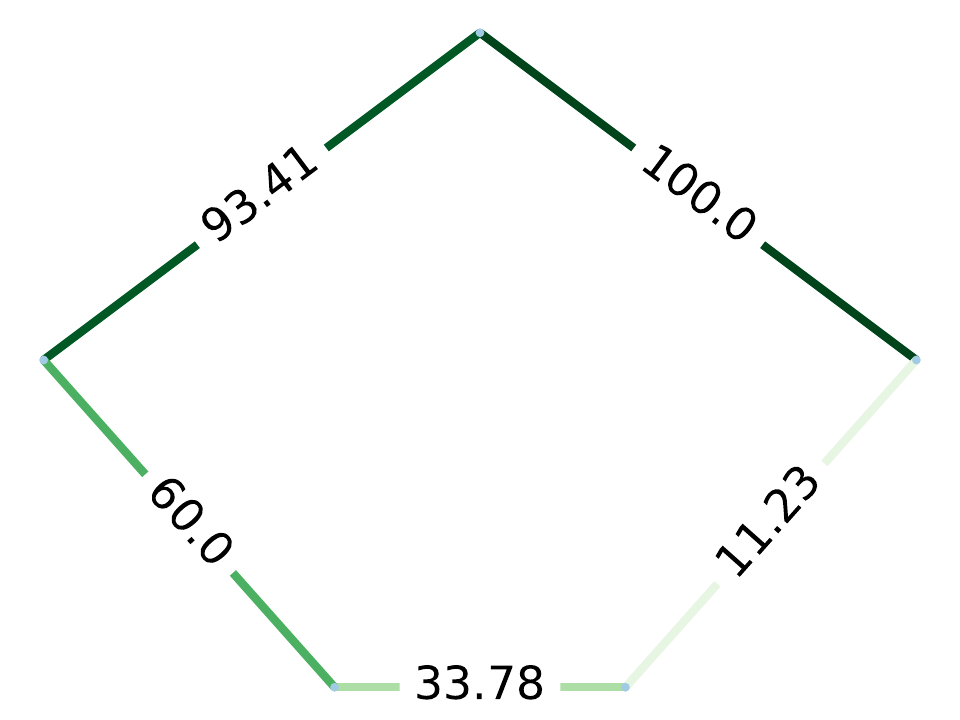} 
\caption{Maximum}
\end{subfigure}
\caption{Unlabelled setting CP interval of level $95\%$. To visualize the set, we plot the extreme elements. The alignment procedure is able to capture the graph topology both for the estimator and the interval on each edge.}
\label{fig:unlabelled_pent_interval}
\end{figure}
In Figures \ref{fig:unlabelled_pent_interval}, we show the Fréchet mean of the simulated dataset and the corresponding $95\%$ intervals. The exact values of the intervals are visualized on top of each edge (the minimum, the Fréchet mean, the maximum). The example shows the capability of the CP intervals to capture both the topology and the attributes when working with an estimation procedure in the unlabelled setting (e.g. the mean).

\section{Empirical Illustration}
\label{sec:Application}
\begin{figure}
    \centering
    \includegraphics[width=\textwidth]{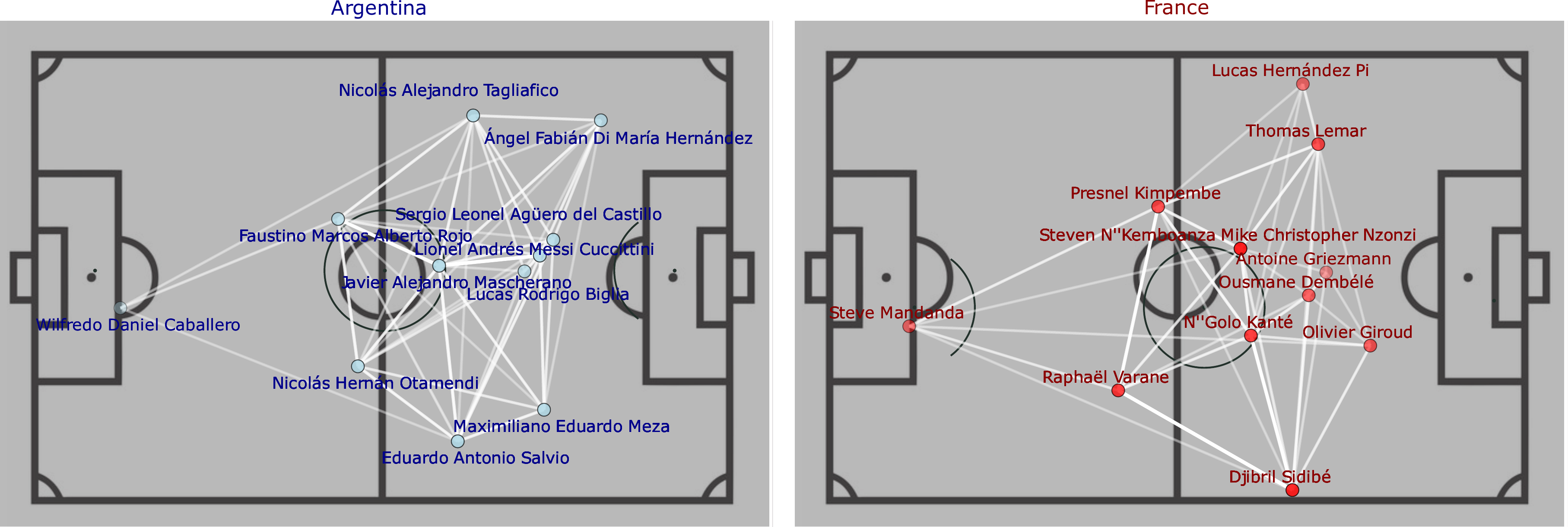}
    \caption{Argentina and France player passing networks during two games. Each node correspond to a player, the edges encode the number of passages. The wider the edge the higher the number of passages between the players.}
    \label{fig:croatiafrance}
\end{figure}

The use of quantitative analytics for performance evaluation has become increasingly popular in professional sports over recent years.
Among the many disciplines in which advanced statistical methods have been used \citep{bunker2019machine, cervone2014pointwise, rudrapal2020deep}, a particular role is played by the analysis of the passing networks of football players. 
The passing network refers to the number of times two players pass the ball to each other during a match. Such analysis can provide insight into team performance, strategy, and ultimately, the final score. 
Various studies have utilized passing network analysis to understand these factors, such as \cite{cotta2013network,clemente2015general,calissano2022graph}. To provide a concrete example of the potential of passing network analysis, we examined the football players passing network (PPN) in the 2018 FIFA World Cup, using an open source database \citep{statsbomb}. The dataset contains observations about the $64$ matches played from $14/06/2018$ to $15/07/2018$ between $32$ teams. In Figure \ref{fig:croatiafrance}, we show the PPN of the final match between the two different teams: Argentina and France. The networks are preprocessed as undirected, focusing on the number of passages between players rather than the direction of the passages. To perform an analysis for the whole championship, PPNs of different teams are compared, requiring an unlabelled approach to the data. As briefly discussed in the Introduction, the comparison between two different teams is possible only if a matching between the players is estimated based on their role in the PPNs. 
We divide teams into high-, low-, and medium-performance categories based on the difference between the number of goals scored and conceded during the match (i.e., $\Delta_{goal}>1$, $\Delta_{goal}< -1$, and $-1 \leq \Delta_{goal} \leq 1$, respectively). For each subset, we use the Fréchet mean as the $\mathcal{A}$ estimator.
%This choice is supported by looking at championship in which the difference between goals scored and the team performance in the tournamnet difference for a team in a World Cup tournament is the 2014 World Cup in Brazil. Portugal did defeat Ghana 2-1 in their final group stage match in the 2014 World Cup, but their goal difference was still not enough to see them through to the knockout rounds. Meanwhile, the United States team was solid enough to concede only a single goal in the defeat against Germany. The United States' overall 0-goal difference was enough to see them through to the knockout rounds as runners-up in their group, while Portugal was eliminated due to their inferior goal difference of -3. This example demonstrates how the difference between goals scored and conceded can have a significant impact on a team's fortunes in a World Cup tournament.

\begin{figure}
    \centering
    \includegraphics[width=\textwidth]{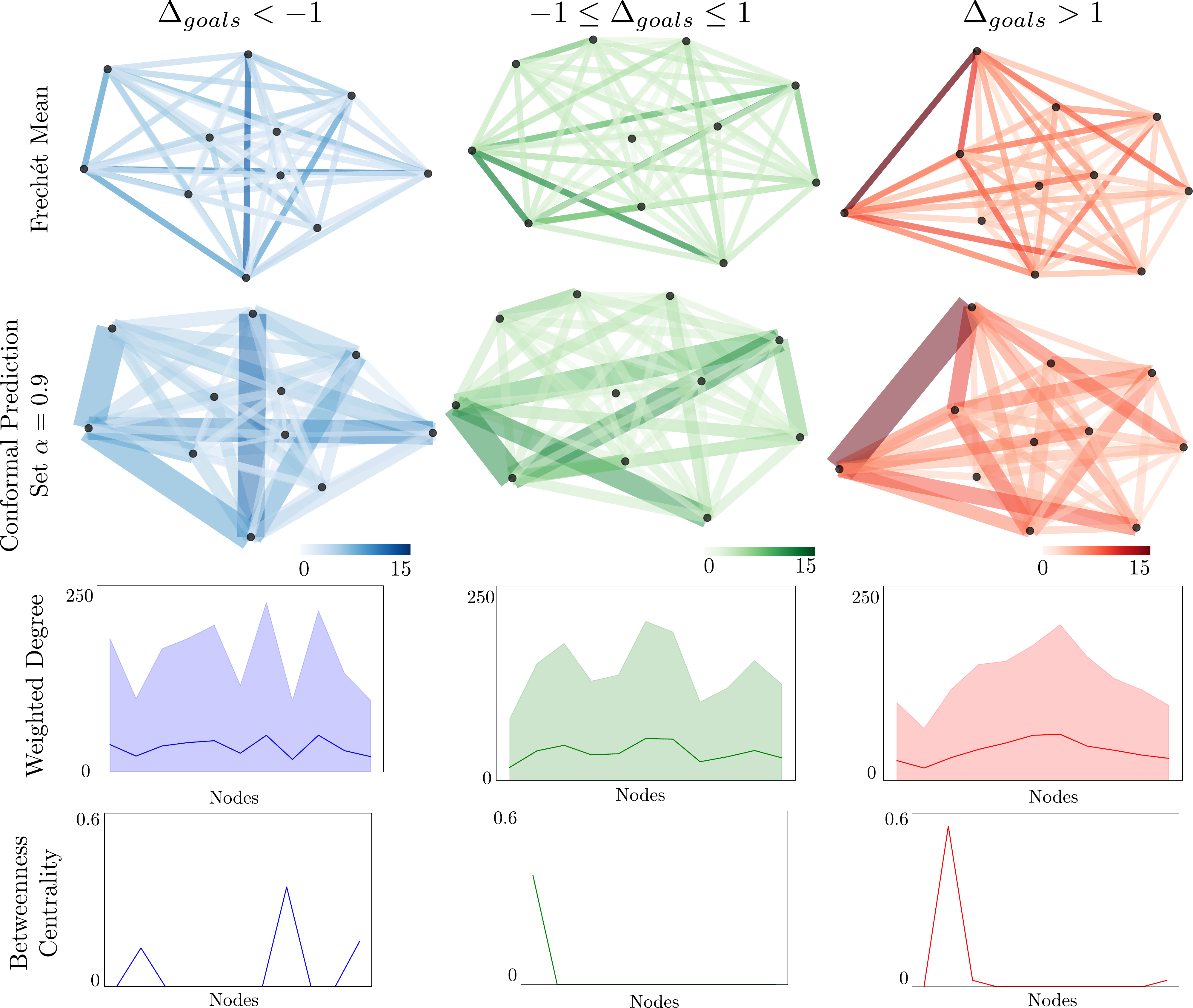}
    \caption{Each column represent a performance level of the team: Low, Medium, High Performance. For each category we represent: PPN Estimator (Fréchet mean), CP set with $\alpha=0.9$ and two indicators (Weighted Degree distribution and the Betweenness Centrality index). The weighted degree is computed for both the mean (line) and the interval (min is set to zero and max is represented as the top line).}
    \label{fig:conf_foot}
\end{figure}

In Figure \ref{fig:conf_foot}, we visualize the Fréchet mean, the CP set with $\alpha=0.9$, and two graph indicators to facilitate the interpretation and the comparison between the groups. In the Fréchet mean plot, we display the edges attributes using the color intensity. In the CP sets plots, we also display the edges attributes using the color intensity, but we added the thickness of the edge to represent the length of the CP set. The CP sets help the interpretation of the results thanks to their component-wise property. In most cases, it can be noticed that the interval on the edges is wider when the number of passages is higher. About the indicators, the degree distribution and the betweenness centrality of the nodes are different across the three groups. Degree distribution refers to the distribution of the number of passes that each player has in the network. It is a measure of the activity of each player: nodes with high degrees are more connected and may be considered more influential in shaping the flow of the ball in the network. Betweenness centrality measures the extent to which a node lies on the shortest paths between other nodes in the network. It is a measure of the importance of a player in controlling the flow of the ball in the network. Players with high betweenness centrality are often involved in a large number of passes between different groups of players, indicating that they play a key role in linking different parts of the team.

Our analysis revealed distinct characteristics for teams with varying levels of performance. By looking at the average graph, high-performance teams exhibited a structured graph topology: see how certain nodes play a unique role in the team comparing to more diffuse topology in low-performance group. They also display a high number of players with a high degree. In addition, the passing networks of high-performance teams appear to be more compact, with players passing the ball to each other more frequently and forming more connections between each other. In contrast, low-performance teams had more players with a lower degree and higher betweenness centrality, indicating a more centralized passing network where fewer players were involved in passing the ball. The topology is also diffused and less structured as shown in the mean plot. However, these players played a more critical role in controlling the flow of the ball. Medium-performance teams had a mix of players with high-degree and high-betweenness centrality, which led to a more diverse passing network where some players were more active in passing, while others played a more central role in controlling the flow of the ball.

The insight gained from the analysis of the passing networks can bring considerable value to a football coach, and football experts in general. It provides a more nuanced understanding of the team's strengths and weaknesses. By examining which players are involved in passing networks during high and low-performing matches, a coach can identify the areas in which the team needs improvement. Moreover, analysis of the passing networks can also help to identify players who may be underutilized in the team's tactics. Coaches can then adjust their tactics to incorporate these players more effectively and maximize their contribution to the team's success.

\section{Discussion and Conclusions}
\label{sec:conclusions}
The issue of predicting with uncertainty complex statistical units is a key research topic in modern statistics, from both a theoretical and applied perspective. Only a few works have been proposed and focus on set forecasting techniques based on either distributional assumptions or heavy computational methods. Both these approaches end in producing forecasts for well-behaved objects, for which the embedding is Euclidean or mildly non-Euclidean. In this work, we address the problem by proposing a model-free and computationally efficient forecasting method, based on CP, for two types of statistical objects of increasing complexity: population of labelled and unlabelled graphs with scalar attributes on nodes and edges. We first describe the component-wise CP sets for labelled graphs - represented as adjacency matrices in the Euclidean space - and then we extended the CP sets to the graph space quotient space - a natural embedding for unlabelled graphs.

A first extension of the proposed framework is to consider population of graphs with more complex Euclidean (and potentially even non-Euclidean) attributes. An example of such data objects are graphs with shapes encoded in the edge attributes \citep{guo2022statistical}. A second extension of the current work regards the embedding strategy. The graph space choice is a convenient embedding as it allows a Euclidean representation of the same dimension for the adjacency matrices. However, it treats both absent elements and existing elements having zero attributes as equal and it might not be suited when the graph represents a planar shape with holes - e.g. a skeleton of a complex shape. As a further development, the CP strategy can then be extended to other embedding contexts such as the one proposed in \cite{chowdhury2019gromov}. Aside from graphs, the current methods can be extended to other data embedded in quotient spaces such as images or shapes (see Chapter 2 and Chapter 9 in \cite{marron_object_2021} for other examples). Last but not least, different the prediction models can be tested. The current framework can be easily extended to other prediction methods, offering a non-parametric model-free strategy to obtain prediction intervals for complex objects.
% The labelled case is derived from a novel method recently proposed in the literature to formulate computationally-efficient prediction sets for functional data. The extension to unlabelled graphs is based on the graph space setting which is able to describe all the problems where the attributes measure the intensity of the edges (e.g. the intensity of social interaction, the intensity of a chemical bound). This is because the zero attribute corresponds to the absence of an edge or a node within the framework. When the attributes describe other features, the presence of zeros should be treated as a multi-modal distribution with a density concentrated in zero. This should be taken into account when building a prediction interval. 

\section*{Acknowledgements} 
The research is partially funded by the corporate social responsibility program of Politecnico di Milano through the PoliSocialAward 2019 winning project Safari Njema. We thank the anonymous reviewers who helped us improving the quality of the manuscript.
\bibliographystyle{elsarticle-harv}
\bibliography{sample,references}

\begin{comment}

\newpage
\section*{Appendix 1}
CP intervals for a different number of observations: $|\mathcal{I}_1|+|\mathcal{I}_2|=30,130,230$ and calibration always to show the discrete possible level of $\alpha$. From the plots, it is clear how the line tents to the theoretical line
\begin{figure}[hbpt!]
\centering
\begin{subfigure}{0.3\textwidth}
\includegraphics[width=\textwidth]{Figures/Simulations/Conformal Prediction Intervals30train_calibration.pdf}
\end{subfigure}
\begin{subfigure}{0.3\textwidth}
\includegraphics[width=\textwidth]{Figures/Simulations/Conformal Prediction Intervals130train_calibration.pdf}
\end{subfigure}
\begin{subfigure}{0.3\textwidth}
\includegraphics[width=\textwidth]{Figures/Simulations/Conformal Prediction Intervals230train_calibration.pdf}
\end{subfigure}
\end{figure}
\end{comment}
\end{document}